\RequirePackage{lineno}\newdimen\linenumbersep\linenumbersep=2pt
\documentclass[sort&compress,aps,prd,5p,superscriptaddress,showpacs,showkeys]{elsarticle}
\usepackage{amsmath,amssymb,color,graphicx,gensymb,bm,braket,siunitx,url,subcaption,soul}
\usepackage[toc,page]{appendix}

\newcommand{\1}[1]{\, \mathrm{#1}}
\newcommand{\n}[1]{\mathrm{#1}}
\begin{document}

\title{Characterization of a Deuterium-Deuterium Plasma Fusion Neutron Generator}

\author[purdue]{R.F.~Lang}
\author[purdue]{J.~Pienaar}
\author[nikhef]{E.~Hogenbirk}
\author[purdue]{D.~Masson}
\author[ptb]{R.~Nolte}
\author[ptb]{A.~Zimbal}
\author[ptb]{S.~R\"ottger}
\author[nyuab]{M.L.~Benabderrahmane}
\author[lngs]{G.~Bruno}
\address[purdue]{Department of Physics and Astronomy, Purdue University, West Lafayette, IN, USA}
\address[ptb]{Physikalisch-Technische Bundesanstalt, Braunschweig, Germany}
\address[nikhef]{Nikhef and the University of Amsterdam, Science Park, Amsterdam, The Netherlands}
\address[lngs]{INFN Laboratori Nazionali del Gran Sasso, Assergi, Italy}
\address[nyuab]{New York University Abu Dhabi, Abu Dhabi, UAE}

\begin{abstract}
We characterize the neutron output of a deuterium-deuterium plasma fusion neutron generator, model 35-DD-W-S, manufactured by NSD/Gradel-Fusion. The measured energy spectrum is found to be dominated by neutron peaks at \SI{2.2}{MeV} and \SI{2.7}{eV}. A detailed GEANT4 simulation accurately reproduces the measured energy spectrum and confirms our understanding of the fusion process in this generator. Additionally, a contribution of $14.1\1{MeV}$ neutrons from deuterium-tritium fusion is found at a level of~$3.5\%$, from tritium produced in previous deuterium-deuterium reactions. We have measured both the absolute neutron flux as well as its relative variation on the operational parameters of the generator. We find the flux to be proportional to voltage $V^{3.32 \pm 0.14}$ and current $I^{0.97 \pm 0.01}$. Further, we have measured the angular dependence of the neutron emission with respect to the polar angle. We conclude that it is well described by isotropic production of neutrons within the cathode field cage.
\end{abstract}

\maketitle

\section{Introduction}

Neutron generators are a convenient, commercially available source of neutrons widely used in science and engineering. They can easily achieve a tuneable neutron flux of $10^6\1{n/s}$ with some generators operating above the $10^{10}\1{n/s}$ range, they pose no or only minimal safety concerns when turned off, and they are available in a variety of configurations. The latest advances in the field of compact sealed-tube neutron generators toward the development of smaller, lighter and less expensive systems further extend their applicability.

Two main reactions are exploited in such generators: deuterium-tritium fusion yielding \SI{14.1}{MeV} neutrons, and deuterium-deuterium fusion yielding \SI{2.45}{MeV} neutrons in the center-of-mass frame. Two operating principles are commonly employed to induce fusion. One is to accelerate a beam of deuterium ions onto a solid state target which contains either deuterium or tritium. Another principle is the fusion of ions in a plasma in the presence of a high voltage potential. Indeed, there are detailed discussions of the characteristics of deuterium-tritium generators~\cite{guillame1971}, deuterium-deuterium generators~\cite{Miley1997, Miley1999, Miley2000} as well as neutron generators in general~\cite{CRC,chernikova}. However, to our knowledge, the measurements reported here represent the first complete characterization of a deuterium-deuterium plasma fusion generator, including the determination of absolute neutron yield, neutron energy spectrum and emission anisotropy.

\section{Setup} \label{sec_setup}

\subsection{Neutron Generator}

The neutron generator characterized in this work is a model 35-DD-W-S deuterium-deuterium plasma fusion neutron generator manufactured by NSD/Gradel-Fusion. This generator produces $2.45\1{MeV}$ neutrons based on the fusion of deuterium 

\begin{equation}
\rm{{}^2D + {}^2D } \rightarrow \rm{{}^3He~}(\SI{0.82}{MeV}) + \rm{n~} (\SI{2.45}{MeV}) {\rm . } \label{eqn:main_reaction}
\end{equation}

The generator is capable of delivering neutron fluxes up to $10^7\1{n/s}$. Given our particular application of this generator in the field of direct dark matter detectors~\cite{xe1t:april2016}, it was modified to enable stable operation even at fluxes as \textit{low} as $10\1{n/s}$. The neutron generator has a cylindrical shape with a length of $940\1{mm}$ and a diameter of $138\1{mm}$. It has a standalone high voltage power supply module, a slow control program to monitor system parameters, and a water cooling loop.

The working principle of the neutron generator is based on inertial electrostatic confinement (IEC). The generator has a fusion chamber filled with deuterium gas. The deuterium pressure is reduced to a level that allows for plasma ignition by glow discharge (Paschen's law). The primary source of neutrons is considered to be the plasma in the volume surrounded by the cathode. Deuterium gas in the fusion chamber is ionized and the resulting ions are accelerated toward the cathode field cage. Once inside the field cage, the ionized gas is confined by applying a high voltage ranging between $10-100\1{kV}$. When the necessary conditions to overcome the Coulomb barrier are met, fusion occurs, emitting approximately mono-energetic $2.45\1{MeV}$ neutrons. 

\subsection{Liquid Scintillators}

We use two liquid scintillator detectors, a 3"$ \times$3" EJ301 cell and a 2"$\times$2" NE213 cell, to measure the neutron energy spectrum and relative flux. The scintillator used in these detectors is identical, apart from the manufacturer \cite{ej_datasheet}. Liquid scintillators are very popular for fast neutron detection as they can easily be shaped into the desired size and geometry of a given application. The process of elastic scattering by neutrons off the protons found in the hydrocarbon molecules produces prompt scintillation that offers excellent timing performance. Such nuclear recoils exhibit greater ionization density rates than electronic recoils that are induced by various backgrounds. Consequently, the ionization tracks of nuclear recoils produce higher yields of delayed fluorescence, resulting in scintillation pulses that decay more slowly than those of electronic recoils of comparable energy. The different pulse shapes that arise from electronic and nuclear recoils in liquid scintillators can thus be exploited using pulse shape discrimination methods~\cite{brooks1959, kuchnir1968, perkins1979, lang2016}. Additionally, if the detector response to neutrons of specific energies is well known, these detectors can be used to reconstruct the energy spectrum of the incident neutron flux~\cite{klein2002,verbinski1968}.

The EJ301 detector cell was optically coupled to a ETEL 9821KB photomultiplier (PMT) which was operated at a voltage of 1700 V. The anode signal of the PMT was acquired using a CAEN DT5751 digitizer, which samples at 1 GHz with 10 bit resolution and has a 1 V dynamic range. The NE213 detector cell was coupled to an XP2020 PMT via a short lucite light guide. The PMT was operated at a voltage of -1950~V. Standard nuclear electronics modules were used for analogue signal processing. A signal proportional to the total amount of scintillation light produced in the detector cell (pulse height) was derived from the ninth PMT dynode. A second signal related to the decay time of the light pulse (pulse shape) was derived from the PMT anode using the zero-crossing technique. The two signals were digitized using pulse-height sensitive ADCs and a PC based multi-parameter data acquisition system. Deuterium-deuterium IEC fusion devices are known to produce bremsstrahlung in addition to neutrons~\cite{Luo2010}. Therefore we operated the liquid scintillator detectors with lead shielding.

The gain of the PMT in the NE213 detector was constantly measured and adjusted using a feedback loop. This was constructed with an integrated LED, that gave constant light pulses at a rate of \SI{65}{Hz}, and a voltage added to the bias voltage based on the PMT's response to the LED pulse. This setup allowed for a constant gain during the long measurements required for determining the energy spectrum (section~\ref{sec:en_spectrum}).

\subsection{De Pangher Long Counter}
We use a De Pangher Long Counter to measure the absolute neutron flux. This detector is made of concentric layers of polyethylene and borated polyethylene. The borated polyethylene shields the detector from neutrons entering the detector through its side wall, whereas incident neutrons from the source, which enter through the planar surface, are thermalized by the polyethylene. At the center of the Long Counter is a tube filled with BF$_3$ that is enriched in $^{11}$B. Thermalized neutrons that are scattered into this volume can be captured, producing an alpha through the $\mathrm{^{11}B(n, \alpha)^{7}Li}$ reaction. The detection of the emitted alpha particle produces a unique signal in the detector that is used to tag neutrons.

The Long Counter has the ability to measure neutron fluxes with a response that is almost independent of neutron energy in the range from a few keV to almost \SI{10}{MeV}. Their high sensitivity allows them to be used to measure low neutron fluxes. Additionally, they are directional, which is necessary to control the impact of in-scattered neutrons~\cite{nolte2011}.

\subsection{Experimental Setup}

\begin{figure}[!htb]
\centering
\includegraphics[width = \columnwidth, clip=true, trim =0 0 0 0]{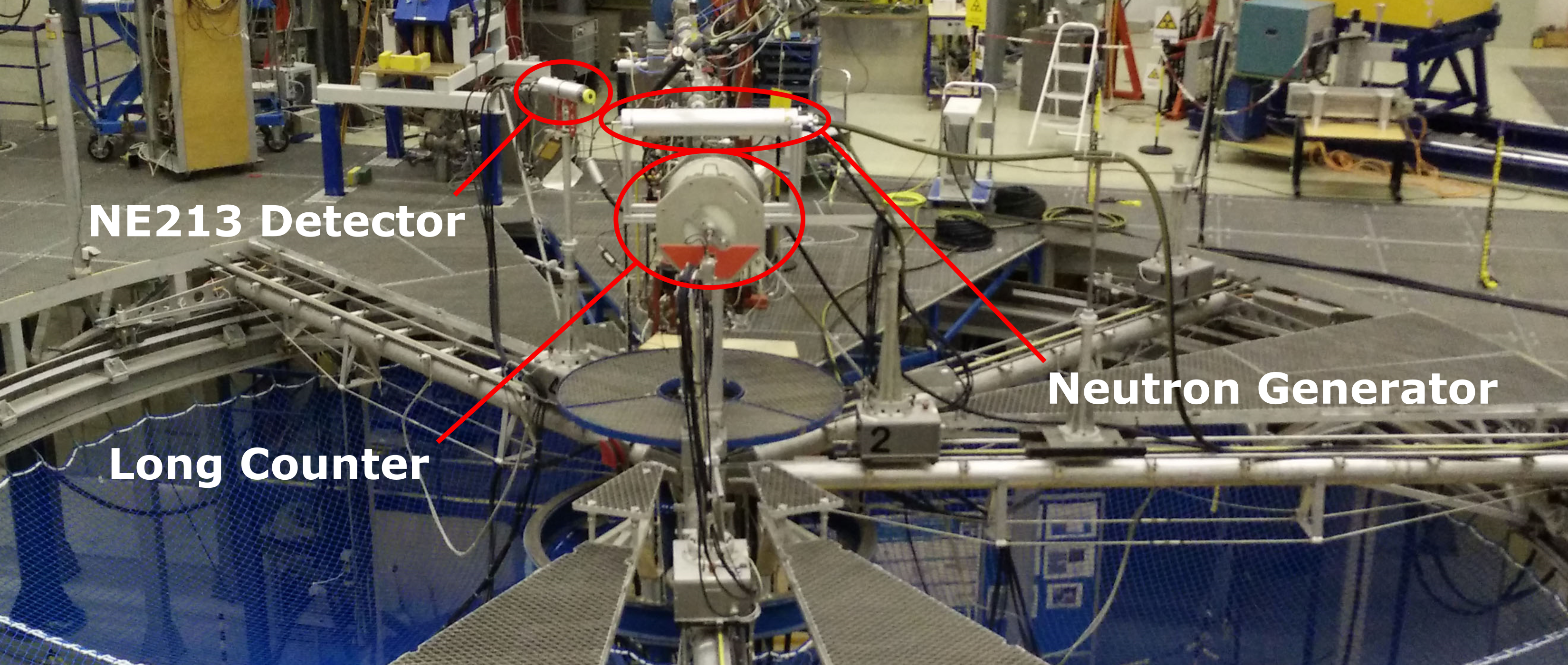}
\caption{The experimental setup at PTB. The neutron generator is located at the center of the circle with the Long Counter at a radial distance of \SI{1569}{mm}. On the left hand side, the NE213 detector can be seen at a distance of \SI{2069}{mm} and at polar angle \SI{230}{\degree}.}
\label{fig:ptb_setup}
\end{figure}

\begin{figure}[!htb]
\centering
\includegraphics[width = 0.9\columnwidth, clip=True, trim= 0  0 0 0]{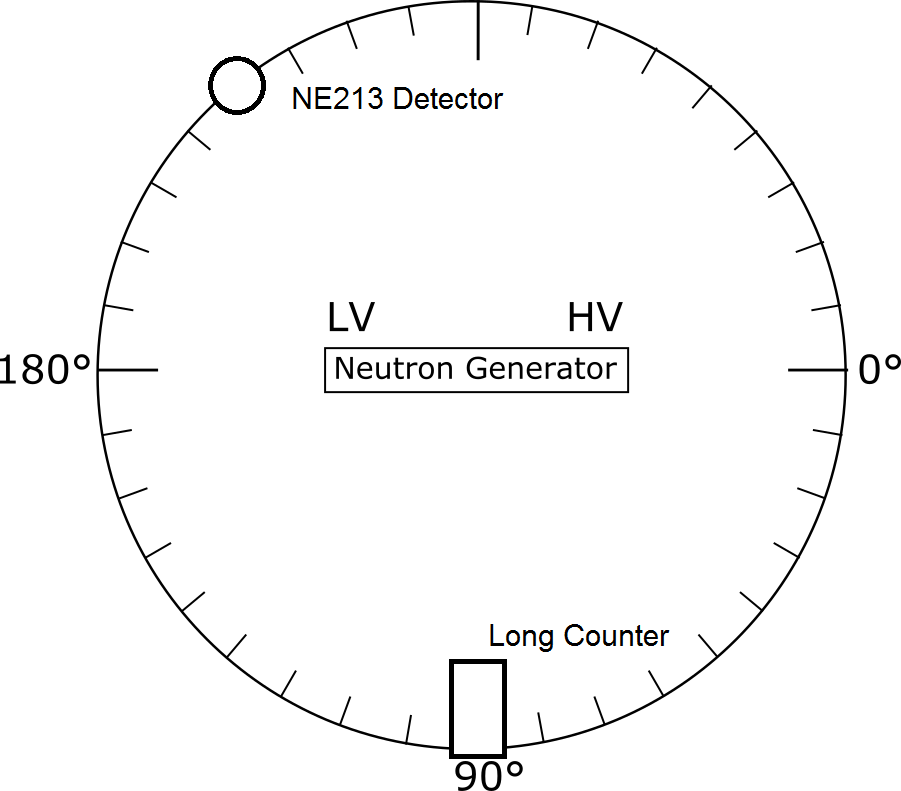} % left lower right upper
\caption{The coordinate system used at PTB. The schematic shows the neutron generator from above, with its high voltage  (HV) end taken to be at \SI{0}{\degree} polar angle and its low voltage (LV) end taken to be \SI{180}{\degree} polar angle. Also shown are the Long Counter at \SI{90}{\degree} and the NE213 detector at \SI{230}{\degree}, as the detectors were placed when Figure~\ref{fig:ptb_setup} was taken.}
\label{fig:ptb_coordinates}
\end{figure}

Measurements of the neutron flux were performed both at Purdue University and at the Physikalisch-Technische Bundesanstalt (PTB). Figure~\ref{fig:ptb_setup} shows the experimental setup at PTB. The neutron generator was placed with its horizontal axis at the center of the experimental facility. We define the polar angle such that the direction perpendicular to the axis of the neutron generator is \SI{90}{\degree}, as shown in Fig.~\ref{fig:ptb_coordinates}. All distances in this section are measured from the center of the neutron generator to the front face of the detector in question.

To measure the neutron energy spectrum, a NE213 detector was placed at a radial distance of \SI{865}{mm} from the generator, facing it at a \SI{90}{\degree} polar angle. These measurements were taken for a duration of \SI{6.0}{hours} while the neutron generator was operated at \SI{50}{kV} and \SI{2.5}{mA}. To be able to perform background discrimination and subtraction, data were also collected for \SI{21.1}{hours} while the neutron generator was off. 

Additionally, data were taken with a shadow cone placed between the neutron generator and the Long Counter, where the Long Counter was placed at two different radii. These measurements were used to verify that the Monte Carlo simulations of the experimental setup correctly accounts for the number of neutrons scattered off the air into the Long Counter.

For the measurements of the angular emission (section~\ref{sec:angular}), the Long Counter was placed on a radial arm at a distance of \SI{1569}{mm} from the neutron generator but at various polar angles. A 3''$\times$3'' EJ301 detector was placed at a distance of \SI{2000}{mm} at a fixed angle of \SI{230}{\degree} in order to have a permanent measurement of the stability of the generator during the angular scan. 

The functional dependence of the neutron flux on the applied voltage and current were determined using three EJ301 liquid scintillator cells at Purdue in various orientations. The response of these detectors to~\SI{2.45}{MeV} neutrons has previously been characterized~\cite{lang2016}. Furthermore, the functional dependence and absolute flux were measured at PTB with the Long Counter at a radial distance of \SI{1569}{mm} and a polar angle of \SI{90}{\degree}.

\section{Monte Carlo Simulation}

We have developed a detailed Monte Carlo simulation of the experimental setup in order to assist with the interpretation of the obtained data. The simulation was developed using the GEANT4 toolkit ~\cite{Agostinelli2003250}.  Technical drawings of the neutron generator and its interior, as well as the three different neutron detectors, were used to create a complete description of the major components that can produce significant scattering of neutrons.

\subsection{Physics List in GEANT4}

This simulation made use of version \texttt{9.4-patch02} of the GEANT4 toolkit. Since the energy of the neutrons of interest is below \SI{20}{MeV}, we use the \texttt{High Precision} physics list, with \texttt{G4NDL 3.14}. This list contains cross-sections down to thermal energies in order to accurately describe the elastic, inelastic and capture processes of neutrons in the Long Counter. The radioactive $\alpha$, $\beta^{+}$, $\beta^{-}$ or electron capture decays are simulated using \texttt{G4RadioactiveDecay}. Information about the half-lives, nuclear level structure, decay branching ratios, and the energies of decay processes are taken from the Evaluated Nuclear Structure Data Files (ENSDF)~\cite{Bhat1992}.

The tracking of particles in GEANT4 is divided into spatial steps. The length of these steps is automatically set depending on the energy and type of each particle, as well as the material it is propagating through. For each interaction, we record the position, deposited energy, particle type, initial energy of the particle, and the process responsible for the energy loss.

\subsection{Neutron Generator Model}\label{sec:initialneutronspectrum}

The description of the neutron generator in the GEANT4 toolkit is reproduced from technical drawings and information provided by the manufacturer, NSD/Gradel Fusion. The important internal components included in the simulation are the reaction chamber and cathode, high voltage feedthrough, getter pump, and the water cooling system. The deuterium gas conditions inside the fusion chamber were modelled using the known pressure of the deuterium gas. We assumed homogeneous neutron production within the cathode volume, producing an isotropic neutron flux. This assumption is consistent with the measured angular neutron flux (section~\ref{sec:angular}) and the energy spectrum measured at two different polar angles (section~\ref{sec:en_spectrum}).

The energy spectrum used as an input in the GEANT4 simulation was calculated using the known energy-dependent differential cross-section of deuterium-deuterium fusion. The characteristic deuteron energy was described by a Gaussian distribution with a mean of ~\SI{30}{keV} and sigma of ~\SI{3}{keV}. While this is consistent with the applied high voltages of \SI{40}{kV} and \SI{50}{kV}, at which measurements of the energy spectrum were obtained, assumed average kinetic energies between \SI{30}{keV} and \SI{50}{keV} also fit the data. 

The energy of the incident particle in a fusion event is determined by randomly sampling from the aforementioned Gaussian distribution. Angles ($\theta \in [0, \pi]$) were randomly sampled to describe the emission angle of the neutron relative to the direction of the incident particle. The energies of the neutrons were determined from scattering kinematics given $\theta$. Differential production cross-section information for neutrons from deuterium-deuterium fusion on thin targets in the center-of-mass frame were taken from~\cite{LISKIEN1973569}. Using the parametrization suggested therein, the differential production cross section can be described by

\begin{equation}\label{eqn:angular_yield}
\frac{d \sigma}{d \Omega}(\theta) = \frac{d \sigma}{d \Omega}(0^{\circ}) \sum\limits_{i} A_i P_i(\theta)  ,
\end{equation}
where $\frac{d \sigma}{d \Omega}(0^{\circ})$ is the differential production cross section at $0^{\circ}$, and $A_i$ are the recommended Legendre coefficients in the center-of-mass frame for a particle with incident energy $E_d$. Using Equation \ref{eqn:angular_yield}, we calculate the production cross-section of neutrons in the lab frame.

The resulting lab frame neutron energies are used to produce an energy histogram, where each entry is assigned a weighting factor determined by the production cross-section of the fusion event that produced a neutron of that energy. During Monte Carlo studies, neutron energies are produced by randomly sampling from this distribution.

\subsection{Liquid Scintillator Detector Model}

The two liquid scintillators EJ301 and NE213 were modeled as simple cylinders of the appropriate dimensions. The material properties of both liquid scintillators were taken to be those of EJ301~\cite{Eljen}. 

\subsection{Long Counter Model}

The De Pangher Long Counter simulation geometry was also reproduced from technical drawings. The concentric layers of polyethylene and borated polyethylene were implemented, along with the central volume of enriched BF$_3$. In the Monte Carlo calculation, the emission of an $\alpha$-particle following neutron capture on $^{11}$B in the central tube is assumed to represent a neutron detection event.

\begin{figure*}[!htbp]
\centering
\includegraphics[width=1.0\textwidth]{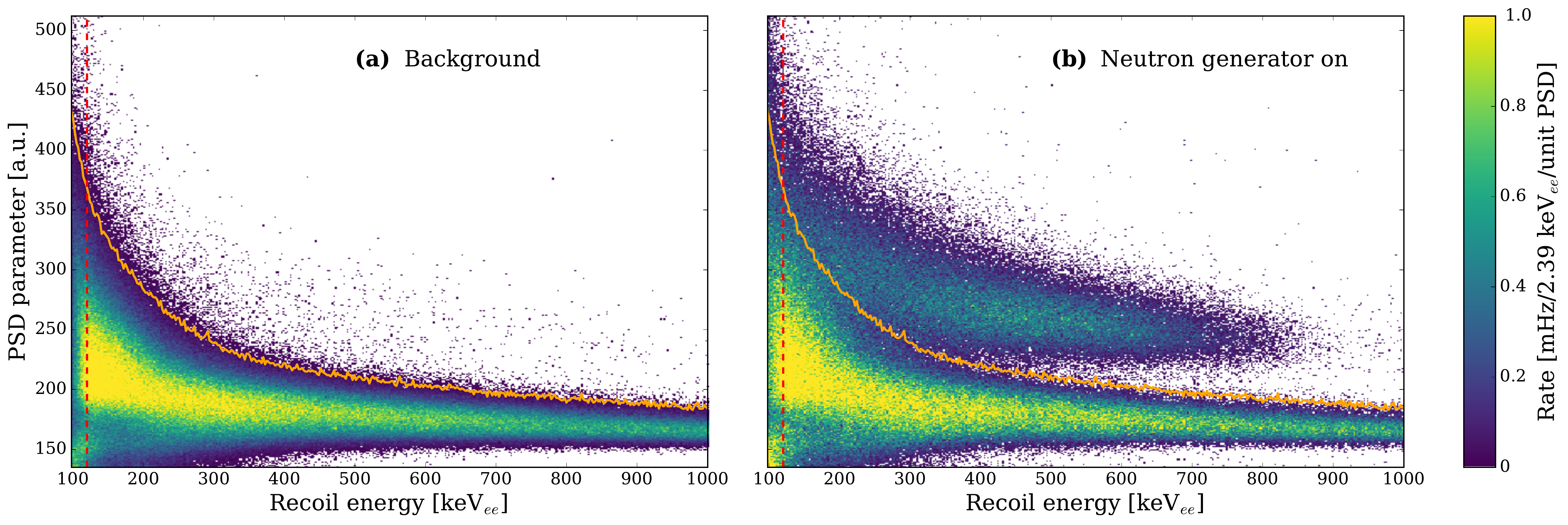}
\caption{{\bf (a)} Data taken with the NE213 scintillator from a 21-hour background run and {\bf (b)} a 6-hour run with the neutron generator turned on, showing the pulse shape discrimination parameter as a function of recoil energy. A clear second neutron-induced band can be seen in the right panel. These recoils are selected using a $99\1{\%}$ background rejection cut in the pulse shape parameter, as indicated by the solid (black) line.
The vertical dashed (red) line indicates the analysis threshold of \SI{120}{keV_{ee}} electron-equivalent energy.}
\label{hist2d}
\end{figure*}

\section{Neutron Energy Spectrum}\label{sec:en_spectrum}

We use the 2"$\times$2" cylindrical NE213 detector to determine the energy spectrum of the neutrons produced by the generator. Since this detector only measures the energy of the recoiling proton, there is no unique measurement of the incident neutron energy. Nevertheless two methods can be used to determine the incident neutron energy spectrum. Both require knowledge of the response function of the scintillator, which is simply a matrix that gives the distribution of the amount of scintillation light for a given incident neutron energy. We use a Monte Carlo-derived response function, which is based on measurements in monoenergetic and broad ns-pulsed neutron fields at PTB. The latter allows one to apply time-of-flight methods to select specific neutron energies \cite{dietze1982, klein2002,zimbal2006}.

If the initial neutron energy spectrum is known, the observed recoil spectrum can  be calculated from the convolution of this response function with the incident spectrum, as we show in section~\ref{sec:convolution}. Alternatively, the observed recoil spectrum can be deconvoluted with the help of the response function in order to extract the incident neutron spectrum, as we do in section~\ref{sec:deconvolution}. The results that we obtain from both methods are in good agreement with each other.

\subsection{Data Selection} \label{sec:data_selection}

The two parameters available for each event in the NE213 scintillator are the pulse height, which increases with the nuclear recoil energy, and the Zero Crossing Method pulse shape discrimination parameter ~\cite{Alexander1961}, which allows one to distinguish nuclear and electronic recoil events. Fig.~\ref{hist2d} shows the histograms for both background data and data taken with the neutron generator turned on. The pulse height is converted to recoil energy (electronic recoil equivalent) by Monte Carlo-matching of a $^{207}$Bi gamma calibration spectrum. For proton recoil events, the nonlinear relation between proton recoil energy and the resulting signal amplitude, known as the light output function, is considered in the neutron response function data \cite{novotny1997}.
A linear energy scale is assumed, which we verified to be correct at six energies ranging up to an electron-equivalent energy of \SI{1546}{keV_{ee}} using the Compton edges of the calibration sources $^{207}$Bi, $^{22}$Na and $^{137}$Cs.

We select neutron-induced events by cutting at the $99\1{\%}$ electronic recoil background rejection line in the pulse shape discrimination parameter space, as indicated in Fig.~\ref{hist2d}. We subtract the background rate in the selected region by computing the number of events that pass the cut in the background data set and using the appropriate scaling for live-time. 
We apply an energy threshold of \SI{120}{keV_{ee}} electron-equivalent energy in order to stay above energies where characteristic X-ray emission from the lead surrounding the NE213 detector starts to dominate. 

\subsection{Cut Acceptance}

\begin{figure}[!htbp]
\begin{center}
\includegraphics[width=1\columnwidth]{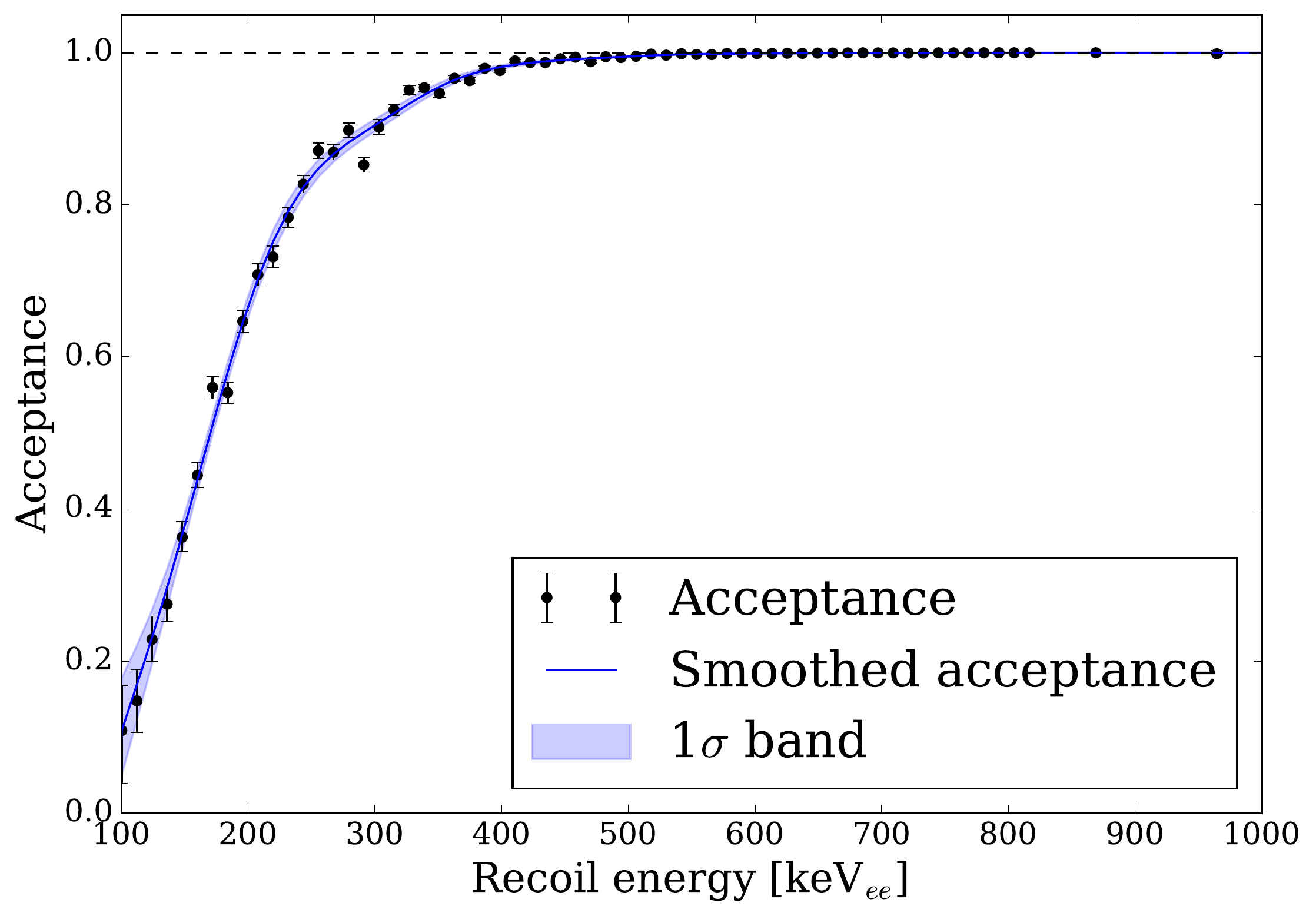}
\caption{The acceptance of the pulse shape cut (shown in Fig.~\ref{hist2d}) as a function of energy. The Gaussian fraction above the pulse shape cut is shown by the (green) data points. The two points at the highest recoil energies are calculated with a wider bin width due to low statistics in the neutron band at these energies. The (blue) solid line is a lowpass-filtered interpolation of those data points. The $1\sigma$ uncertainty band on this acceptance is also shown (shaded blue).}
\label{acceptance}
\end{center}
\end{figure}

At low recoil energies, the pulse shape discrimination becomes less efficient. Due to $99\1{\%}$ background rejection criterion, for lower pulse heights an increasingly bigger fraction of neutron events does not fall in the selected region of neutron events. We calculate the fraction of neutron events passing this cut (the \emph{acceptance}) as a function of recoil energy as follows: for each energy bin, we subtract the live-time normalized background data (Fig.~\ref{hist2d}a) from the neutron generator data (Fig.~\ref{hist2d}b) and fit a Gaussian to the resulting nuclear recoil distribution. The neutron acceptance, shown in Fig.~\ref{acceptance}, is then taken as the area fraction of the Gaussian above the background rejection line. Also shown in the Fig.~\ref{acceptance} is a smoothed interpolation that we use in the calculations that follow. We calculate the uncertainty on the acceptance by changing the fit parameters of the Gaussian within their uncertainty and re-computing the resulting acceptance.

\subsection{Convolution}\label{sec:convolution}

\begin{figure}[!htbp]
\begin{center}
\includegraphics[width=\linewidth]{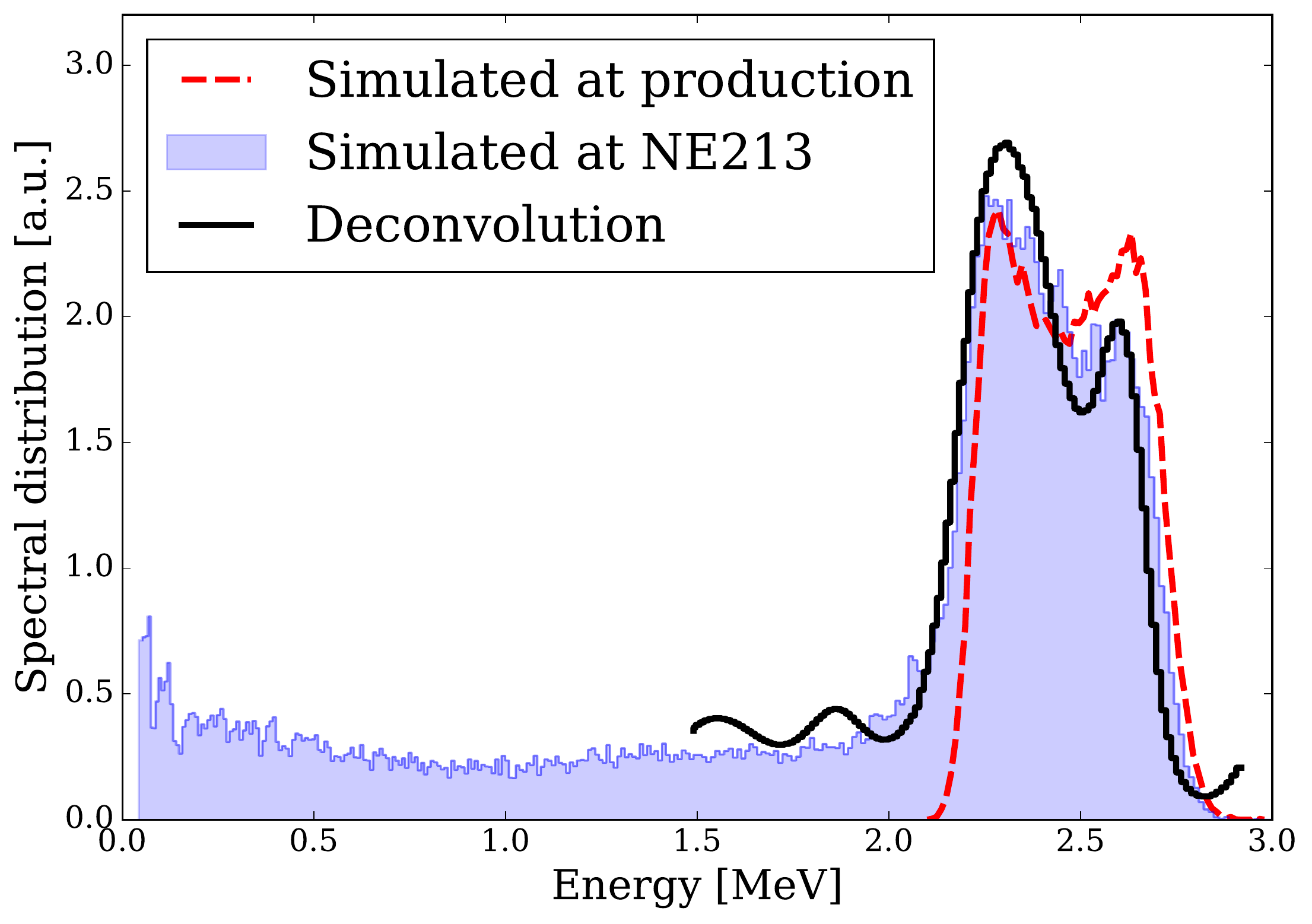}
\caption{Shown are the simulated neutron spectrum in the neutron generator fusion region (red dashed line) and the resulting simulated spectrum at the NE213 detector (blue shaded histogram). Also shown is the result of the deconvolution of our data (black solid line), which is discussed in section~\ref{sec:deconvolution}. All spectra are normalized to the region \SIrange{2.2}{2.7}{MeV} for ease of comparison.}
\label{incident_spectrum}
\end{center}
\end{figure}

The initial neutron energy spectrum in the neutron generator fusion region, calculated as described in section~\ref{sec:initialneutronspectrum}, is shown in Fig.~\ref{incident_spectrum}. As can be seen, a plasma fusion generator as used here does not produce a truly monoenergetic neutron spectrum. Due to the dependence of the fusion cross section on the neutron emission angle relative to the momentum of the incident deuteron, the spectrum shows two peaks at \SI{2.22}{MeV} and \SI{2.72}{MeV} in the lab frame, corresponding to emission angles of \SI{180}{\degree} and \SI{0}{\degree}, respectively.

Neutrons from this spectrum are propagated from the fusion region using GEANT4. The resulting neutron energy spectrum at the NE213 detector position is also shown in Fig.~\ref{incident_spectrum}, where we normalized the spectra to the energy range from \SIrange{2.2}{2.7}{MeV} for ease of comparison. The most significant impact to the neutron energy spectrum comes from scattering in the cooling water that surrounds the neutron generator fusion chamber. This causes the long tail towards lower energies and gives rise to the asymmetric peak structure.

\begin{figure}[!htbp]
\begin{center}
\includegraphics[width=\linewidth]{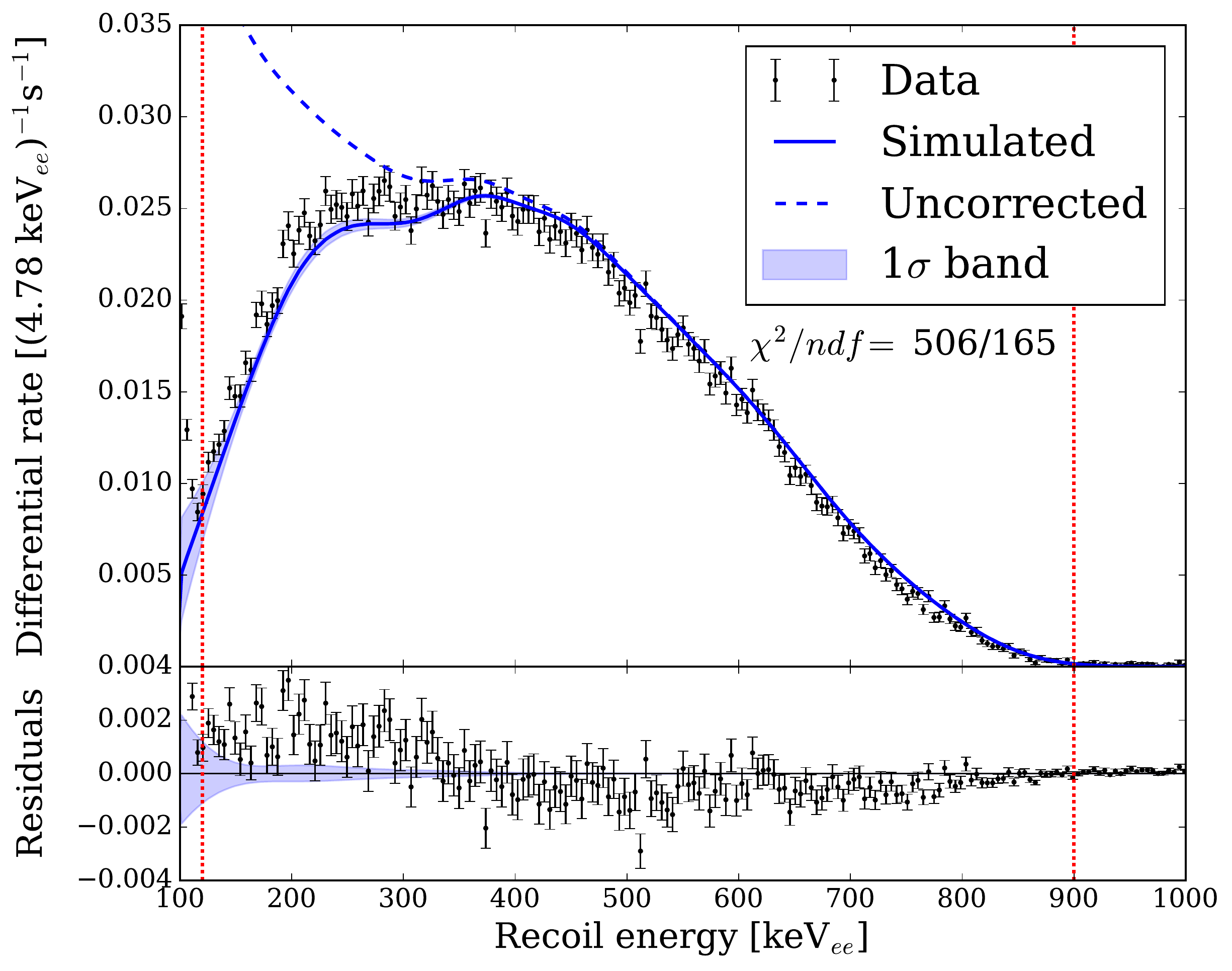}
\caption{The observed background-rejected energy spectrum (green data points) together with the normalized expected distribution from simulation (solid blue line). The energy range considered for the normalization is indicated (red dotted lines). The reduced $\chi^2$ of the fit is 3.07. An indication of the effect of the acceptance correction can also be seen (dashed line). The (light blue) band around the simulated distribution indicates the uncertainties from the acceptance function. The bottom panel shows the residuals between data and simulation after normalization. }
\label{fit_with_acc}
\end{center}
\end{figure}

We calculate the expected pulse height spectrum observed by the NE213 detector from the incident Monte Carlo-derived neutron energy spectrum. To this end, we apply the response function described earlier and use the data-derived acceptance function (Fig.~\ref{acceptance}) to correct for acceptance losses at low pulse height energies. We normalize the expected pulse height spectrum to the observed data using a $\chi^2$-minimization in the energy range $(120-900)\1{keV_{ee}}$ electron-equivalent energy, where the upper limit of the considered energy range is placed at the point where signal and background rates become comparable. The result of this normalization is shown in Fig.~\ref{fit_with_acc} with its $1\sigma$ uncertainty, and the residuals. We find good agreement between the expected and observed distributions, validating the assumed energy spectrum of produced neutrons shown in Fig.~\ref{incident_spectrum}. A closer inspection of the residuals of the minimization indicates that our nuclear recoil acceptance may be slightly underestimated below $350\1{keV_{ee}}$.

In addition to the data selection method outlined in section~\ref{sec:data_selection}, we repeated the analysis with no selection on the pulse shape discrimination parameter. The pulse height spectrum from the neutron generator is in this case computed by subtracting the background spectrum from the spectrum taken with the neutron generator turned on. Since this method requires no acceptance correction, it can be used as a cross-check of the cut-based analysis, however, with larger statistical uncertainty from the increased number of bins. Any gamma radiation caused by neutrons or Bremsstrahlung is ignored in this approach. By comparing the gamma-induced band for the neutron generator and background runs, we estimate that this contributes $\lesssim 10$\% to the number of events between \num{300} and $800\1{keV_{ee}}$. We find that the data selected by this simple background subtraction is consistent with the cut-based analysis if this effect is considered.

\subsection{Deconvolution}\label{sec:deconvolution}
We determine the neutron spectrum through a deconvolution of  the observed nuclear recoil pulse height spectrum. For this analysis, we use the same data selection, background treatment and acceptance correction as in section \ref{sec:convolution}, but restrict the energy range to $350\1{keV_{ee}}-950\1{keV_{ee}}$ electron-equivalent energy.
The main purpose of the deconvolution is an independent confirmation of the expected line shape of the \SI{2.45}{MeV} neutron peak resulting from reaction kinematics and Monte Carlo modeling of the setup. The recorded pulse height events below $350\1{keV_{ee}}$ do not contribute to this information but cause instabilities in the deconvolution process due to non-perfect neutron-gamma separation and systematic limitations in the precise determination of the response function in this energy range.

We use a combination of the GRAVEL \cite{Matzke1994} and MAXED \cite{reginatto2002} deconvolution codes, with GRAVEL providing the starting values that are then used for further refinement using the MAXED code. The allowed neutron energy range for this analysis is \SIrange{1.49}{2.92}{MeV}, corresponding to the light output of the selected pulse height range.

The neutron spectrum obtained from this deconvolution is shown in Fig.~\ref{incident_spectrum}. In agreement with the Monte Carlo simulation, this deconvolved spectrum shows a double-peaked structure around \SI{2.4}{MeV}, rather than a purely monoenergetic neutron emission. At low energies ($<\SI{2.0}{MeV}$), an oscillatory feature appears. However, only a small fraction of the low-energy neutrons can induce a signal above our analysis threshold of \SI{350}{keV_{ee}}. Consequently, small changes in the analysis (such as the considered energy range or data selection criteria) result in significant changes to the spectral shape below \SI{2.0}{MeV}. We thus do not consider this neutron energy range any further. A small contribution is seen at the highest allowed energies ($>\SI{2.8}{MeV}$). As discussed in the next section, we attribute this contribution to the presence of high-energy neutrons from deuterium-tritium fusion.

\subsection{High-Energy Neutrons} \label{he_neutrons}

Upon inspection of the NE213 nuclear recoil data at high ($>1000\1{keV_{ee}}$ electron-equivalent energy) energies, we observed a secondary population at energies above those expected from \SI{2.45}{MeV} deuterium-deuterium neutrons. We attribute these high-energy recoils to deuterium-tritium fusion, produced in the neutron generator as a result of the reactions
\begin{eqnarray}
{}^2\mathrm{D} + {}^2\mathrm{D} \rightarrow \mathrm{p}\, + &\!\!\!\!{}^3\mathrm{T}\!\!\!\!& \label{t_production_reaction} \\
                                              &\!\!\!\!^3\mathrm{T}\!\!\!\!& + {}^2\mathrm{D} \rightarrow {}^4\mathrm{He} + \mathrm{n}(14.1\1{MeV}).
\end{eqnarray}
Reaction~(\ref{t_production_reaction}) is equally likely to occur at $50\1{keV}$ as the main neutron producing reaction~(Eq \ref{eqn:main_reaction}) \cite{huba2013}. Since the deuterium-tritium fusion reaction has a cross section that is more than two orders of magnitude higher than that of deuterium-deuterium fusion, even a small tritium contamination can give a non-negligible amount of \SI{14.1}{MeV} neutrons in the spectrum.

\begin{figure}[!htbp]
\begin{center}
\includegraphics[width=\linewidth]{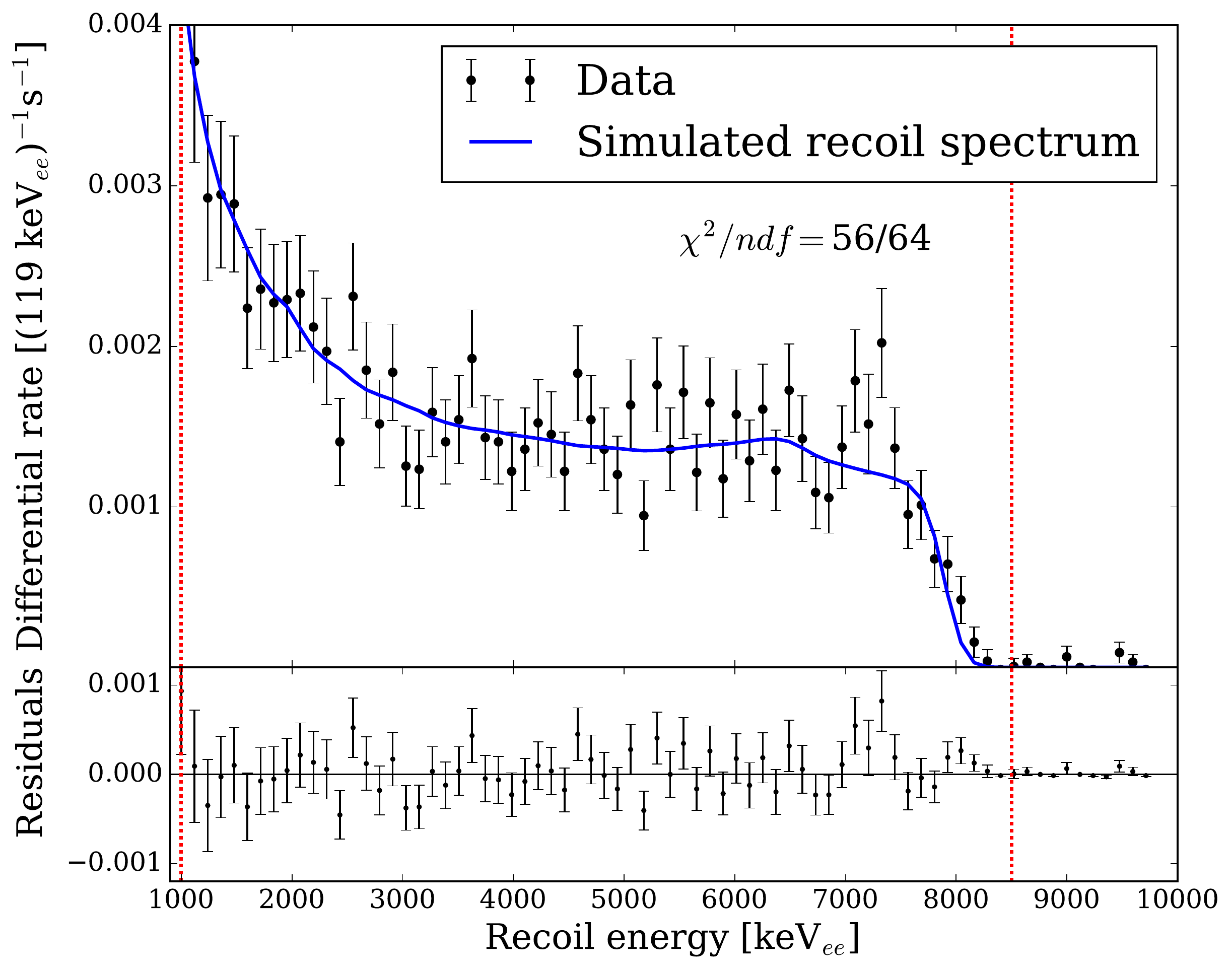}
\caption{The observed spectrum at high recoil energies (green data points) together with the normalized distribution expected from \SI{14.1}{MeV} neutrons from deuterium-tritium fusion (solid blue line). Again, the energy range considered for the normalization is indicated (red dotted lines) and residuals from the normalization are shown in the bottom panel. The reduced $\chi^2$ of the fit is 0.875.}
\label{he_spec}
\end{center}
\end{figure}

To test this hypothesis, we rebin the recoil energy spectrum as shown in Fig.~\ref{he_spec}. Due to the low statistics, the deconvolution code is prone to reconstruction artifacts. We therefore only use the convolution method, with \SI{14.1}{MeV} neutrons as the initial spectrum originating in the neutron generator fusion volume. Again, scatters in surrounding materials were taken into account by propagating the neutron spectrum to the detector in GEANT4. As the separation between the electronic and nuclear recoil bands in the NE213 detector is excellent at these high recoil energies ($>$\SI{1000}{keV}$_{ee}$), the acceptance is taken to be unity. The resulting simulated recoil spectrum is shown in Fig.~\ref{he_spec}, scaled to data. We find an excellent agreement between data and simulation, confirming our hypothesis of deuterium-tritium fusion taking place in the generator.

For a quantitative analysis, we calculate the ratio of the deuterium-tritium neutron flux to the flux of neutrons integrated between \num{2.0} and \SI{2.8}{MeV} in the deconvoluted energy spectrum, taking into account the energy-dependent response of the NE213 detector. We arrive at a total ratio of (\num{5.5} $\pm$ 0.3)\%. This is consistent with the yield expected from tritium produced during the operation of the neutron generator prior to the data taking presented here. To determine the ratio of deuterium-tritium neutrons to the flux of all neutrons from deuterium-deuterium fusion, we calculate the fraction of neutrons produced in the GEANT4 MC simulation with energies below \SI{2.0}{MeV}. In total 37.34\% of the simulated neutrons have incident energies below \SI{2.0}{MeV} when they reach the detector. This is in agreement with the difference in flux measured by the NE213 detector and the Long Counter, discussed at the end of section \ref{sec:flux}. We conclude that the total ratio of deuterium-tritium neutrons is (\num{3.5} $\pm$ 0.2)\%.

\section{Neutron Flux}\label{sec:flux}

\subsection{Relative Dependence}\label{sec:relativeflux}

Measurements were taken at Purdue University to measure the functional dependence of the neutron flux on the operational parameters of the neutron generator, namely the high voltage $V$ and the current $I$ applied to the cathode. High voltages were set to between \SI{30}{kV} and \SI{52}{kV}, and currents were set to between \SI{0.5}{mA} and \SI{1.1}{mA}. Measurements of the high voltage, current, and getter pump temperature were collected in one minute intervals. Three EJ301 organic liquid scintillator detectors were used to measure the neutron flux. Because this experiment was conducted in a small lab, the resulting backscattering of neutrons from the walls prevents us from using this data to determine the absolute neutron flux. This does not however prevent measurement of the relative functional dependence on $V$ and $I$.

A total of 107~datasets with a combined live-time of 55.3~hours were collected with the detectors level with the neutron-producing region of the generator. We have previously studied and improved upon the pulse shape discrimination using the EJ301 detectors. That work allows us to reduce the recoil energy threshold for this analysis to $50\1{keV_{ee}}$ electron-equivalent energy at 99.5\% electronic recoil rejection, using a Laplace transform-based pulse shape discrimination parameter~\cite{lang2016}.

We obtain a rate of nuclear recoils passing this rejection cut and fit a function of the form 
\begin{equation} \label{eq:rate}
F(V,I) = aV^b I^c. 
\end{equation}

As the neutron flux depends strongly on the applied high voltage, which varied during a run by up to several hundred volts, the voltage measurements are averaged together for each run via the expression

\begin{equation}
\langle V \rangle = \left(\frac{1}{n}\sum\limits_{i=0}^n V_i^p \right)^{1/p}
\end{equation}
The exponent $p$ is determined to be 3.33 through an iterative process where a value is chosen, the functional dependence is calculated from the data, and the resulting voltage exponent used to re-average the voltage measurements. This was repeated until the values converged. Typical variations in the applied current were of order $\n{\mu A}$. Since the relative variations are much smaller and the flux depends only weakly on current, no averaging process was required for the current.

To estimate systematic uncertainties in the values of $b$ and $c$, we investigate variations of the selection criteria and also compare the results from the three different detectors. We find the results to be robust against changes in the $50\1{keV_{ee}}$ energy threshold requirement as well as against variations in the exponent $p$ used when averaging high voltage measurements.

\begin{figure}[!htbp]
\includegraphics[width=\columnwidth]{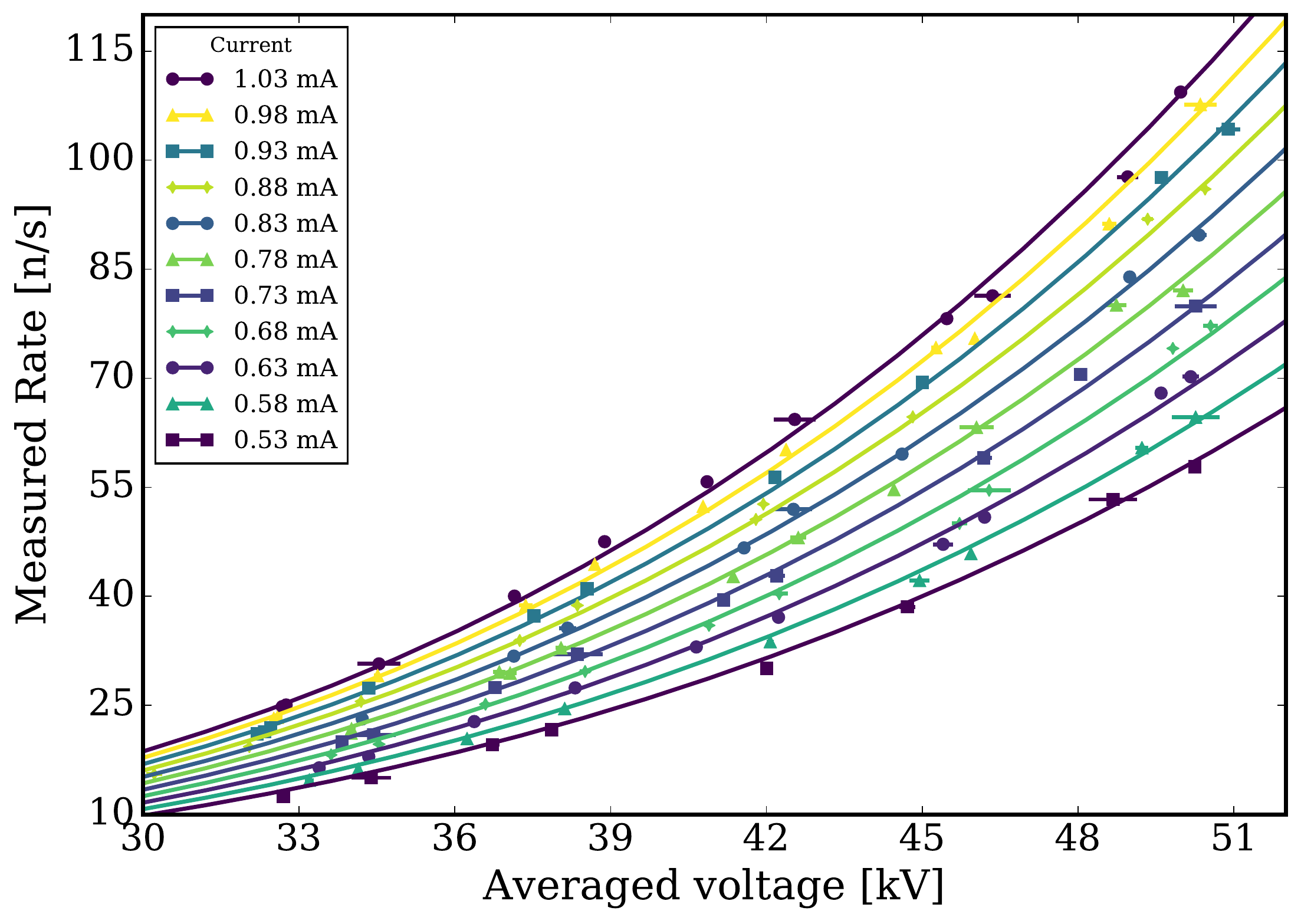}
\caption{Measurements of the relative neutron flux produced by the neutron generator at Purdue University for various values of the operational parameters (cathode high voltage and current). The current at which each measurement was taken is represented by the color of the data point. The flux from the functional fit was found to be proportional to $(\n{Voltage})^{3.32}$ and $(\n{Current})^{0.97}$. The solid lines represent the predicted neutron flux as a function of cathode high voltage, at a fixed current. }\label{fig:fluxandfit}
\end{figure}

Due to scattering of neutrons from the walls, the value found for $a$ is not an accurate measurement of the overall scaling of the function, but we find $b=3.32\pm0.14$ and $c=0.97\pm0.01$. Figure~\ref{fig:fluxandfit} shows the normalized measured neutron rates as a function of cathode high voltage, at fixed current.

\subsection{Absolute Flux}

To measure the absolute neutron flux, a similar measurement was performed at the PTB, where the large experimental hall resulted in significantly less environmental scattering of neutrons. The Long Counter was used as the detector, placed at a distance of $1.569\1{m}$ from the neutron generator, as measured to the front face of the detector. A total of 10~datasets were collected at different high voltage and currents settings over a wider range than measured at Purdue. The same fitting procedure described above was applied, resulting in $b = 3.31 \pm 0.08$, and $c = 1.00 \pm 0.02$, consistent with the Purdue measurements. 

We use the energy-dependent Long Counter ~\cite{npl_report} response, taking into account neutron contributions from both deuterium-deuterium and deuterium-tritium fusion, to correct the measured neutron flux for the space angle covered by the detector. Additionally we correct for the residual environmental scattering from the experimental hall. Furthermore, we also take into account the anisotropic neutron emission presented in the next section. We thus find for the absolute deuterium-deuterium neutron output outside of the neutron generator in the full $4\pi$ space angle (see equation~\ref{eq:rate}) $a = (8.0 \pm 0.6)\times10^{-2}\1{s.^{-1}}$ with $V$ in kV and $I$ in mA. Given the determined values of $a$, $b$ and $c$, the measurements of neutron flux performed here span a range from \num{7.8e3} to \SI{1.16e5}{\second^{-1}}. We expect that the functional dependence of the rate on the operational parameters is valid for neutron fluxes higher than those measured in this work.

As a cross check for consistency, we compare the measured neutron flux in the NE213 detector and the Long Counter.  For the Long Counter we use data collected at a polar angle of \SI{90}{\degree} and an effective radial distance of $1663\1{mm}$. Here we have included the distance to the effective center of the Long Counter, which is $94\1{mm}$ behind the front face of the detector. The neutron generator was set to $2.0\1{mA}$ and $50\1{kV}$, resulting in a measured neutron flux in the Long Counter of $(0.247 \pm 0.022)\1{cm^{-2}.s^{-1}}$. We compare this to data collected with the NE213 detector at an effective radial distance of $892\1{mm}$ (which includes the distance to the effective center of the NE213 detector of $27\1{mm}$) and a polar angle of \SI{90}{\degree}. This yielded a measured neutron flux of $(0.643 \pm 0.064)\1{cm^{-2}.s^{-1}}$ between \num{2.0} and $2.9\1{MeV}$. Taking into consideration the full neutron energy spectrum (discussed in section \ref{he_neutrons}) the total flux is $(1.06 \pm 0.11)\1{cm^{-2}.s^{-1}}$. This value is then corrected for the difference in radial distances between the two detectors and scaled for the different current settings using equation \ref{eq:rate}. The comparable neutron flux from the NE213 detector is thus determined to be $(0.246 \pm 0.025)\1{cm^{-2}.s^{-1}}$.

\section{Angular Emission of Neutrons}\label{sec:angular}

The internal geometry of the neutron generator cylinder is azimuthally symmetric, but not along its axis. We therefore assume that the flux of neutrons is independent of azimuthal angle, and measure the neutron flux as a function of polar angle (compare Fig.~\ref{fig:ptb_coordinates}). An angular scan was performed in steps of \SI{10}{\degree} ranging from \SI{10}{\degree} to \SI{180}{\degree} using the Long Counter. For these measurements, the neutron generator was operated at a high voltage of \SI{50}{kV} and a current of \SI{2}{mA}, and we collected approximately 1000~neutron counts for each measurement. 

\begin{figure}[!htbp]
\centering
\includegraphics[width = \columnwidth]{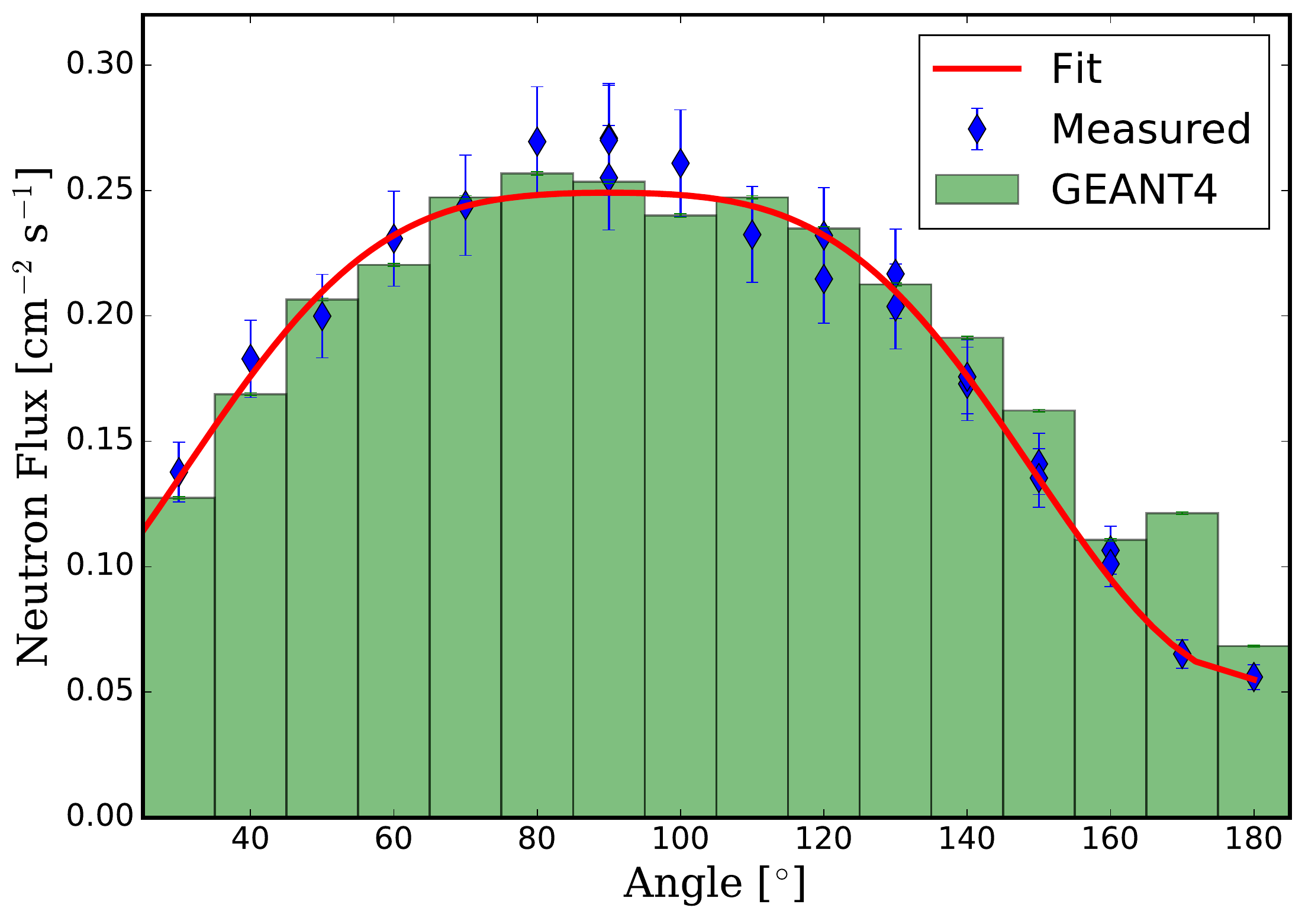}
\caption{Measured neutron flux as function of polar angle. Data taken with the the Long Counter is shown (blue diamonds) together with the angular neutron flux dependence predicted by a detailed GEANT4 simulation of the neutron generator (green bars). A fourth order polynomial fit to the data is shown as well (red line), parametrizing the measured dependence.}
\label{Fig:Angle_Scan_Match}
\end{figure}

Data taken with both the Long Counter and the EJ301 detector at PTB were corrected for the neutron flux that was produced, given the average high voltage and current, as discussed previously. After correction, the measured number of neutrons in the EJ301 detector (which was held at a constant angle throughout) was found to have a standard deviation of $7.5\1{\%}$ across the angular scan measurements. The neutron flux at the location of the Long Counter, shown in Fig.~\ref{Fig:Angle_Scan_Match}, was calculated from the count rate using the energy-dependent response of the Long Counter ~\cite{npl_report}.

A fourth order polynomial, of the form $F(\cos\theta)=A+B\cos^{2}\theta+C\cos^{4}\theta$, is fitted to the Long Counter measurements in order to parametrize the angular dependence of the neutron flux. The resulting fit parameters are $A=(0.237\pm0.006) \1{cm^{-2} s^{-1}}$, $B=(-0.051\pm0.028)\1{cm^{-2} s^{-1}}$ and $C=(-0.130\pm-0.025)\1{cm^{-2} s^{-1}}$.

%F(\theta) = \frac{(36.4 \pm 0.9)}{\sqrt{2 \pi} (52 \pm 1)} e^{\frac{-(\theta+(1.46 \pm 1.5))^2}{2 (52 \pm 1)^2}}
We simulate the expected angular dependence using the detailed neutron generator setup in GEANT4. The Long Counter is placed at the correct distance and angle of each respective measurement. The EJ301 detector is also present in the simulation to account for any scattering off of the EJ301 cell into the Long Counter as the latter is placed closer to the EJ301 cell for large polar angles. At each angle we simulate $10^8$~neutrons distributed homogeneously throughout the cathode and emitted isotropically with the energy spectrum previously calculated. The GEANT4 Monte Carlo data sets are expressed in terms of a neutron flux, using the energy-dependent response function of the Long Counter derived from GEANT4. The resulting simulated angular dependence is also shown in Fig.~\ref{Fig:Angle_Scan_Match}.

The simulation results agree well with data. We conclude that the measured angular emission spectrum is well described by isotropic production of neutrons within the volume of the cathode field cage, with any angular variations being a result of the interior geometry and composition of the neutron generator.

\section{Conclusions}

We have performed the first characterization of the neutron flux produced by a deuterium-deuterium plasma fusion neutron generator. Our interest lies in the application of this generator as a nuclear recoil calibration source for a sensitive dark matter scattering experiment~\cite{xe1t:april2016}. For this application, accurate knowledge of both the energy spectrum and the absolute flux are mandatory.

We found that the energy spectrum is not strictly monoenergetic, but contains two peaks at \SI{2.22}{MeV} and \SI{2.72}{MeV}. These are understood to be caused by the dependence of the neutron energy on the neutron emission angle relative to the momentum of the incident deuteron. Running this generator produces small quantities of tritium that resulted in a measurable flux of \SI{14.1}{MeV} neutrons due to deuterium-tritium fusion in the plasma. These deviations from an ideal, monoenergetic spectrum will have to be taken into account in any measurement using deuterium-deuterium generators that aims to use the energy information of the incident neutron. 

We have also characterized the dependence of the absolute neutron flux on the applied high voltage and current, as well as the dependence of the emitted neutron flux on polar angle. Monte Carlo simulations showed that the angular distribution of the neutron flux is affected by transmission of neutrons through the generator housing. The measured angular distribution is consistent with an isotropic neutron source inside the generator. Taken together, knowledge of this parametrization will allow us to break the degeneracy in calibration between decreasing acceptances of a detector and varying flux output from the neutron source~\cite{Aprile:2013teh}.

We have developed a Monte Carlo simulation of the neutron generator characterized in this work, which accurately predicts the emitted neutron flux. The simulation is able to reproduce both the measured angular emission spectrum and energy of emitted neutrons. Thus we have a predictive model of the behavior of the neutron generator as a nuclear recoil calibration source for other applications.

\section*{Acknowledgements}

We thank the manufacturer NSD/Gradel fusion for many useful discussions and for providing technical drawings of the neutron generator. This work is supported by grants \#PHY-1206061, \#PHY-1209979, \#PHY-1412965 and \#PHY-1650021 from the National Science Foundation (NSF) and carried out under a cooperation agreement between Purdue University and PTB. This work is supported by the research program of the Foundation for Fundamental Research on Matter (FOM), which is part of the Netherlands Organization for Scientific Research (NWO) (grant number FOM VP139). JP~is supported by scholarship \#SFH13071722071 from the National Research Foundation (NRF).

\end{document}